\documentclass[11pt,letterpaper,twoside]{article}
\usepackage{./asp2014}

\aspSuppressVolSlug
\resetcounters

\begin{document}

\title{Exploring eclipsing binaries, triples and higher-order multiple star systems with the SuperWASP archive}
\author{Marcus~E.~Lohr
\affil{Department of Physical Sciences, The Open University, Walton Hall, Milton~Keynes MK7~6AA, UK; \email{Marcus.Lohr@open.ac.uk}}}

\paperauthor{Marcus E. Lohr}{Marcus.Lohr@open.ac.uk}{}{The Open University}{Department of Physical Sciences}{Milton Keynes}{Buckinghamshire}{MK7 6AA}{United Kingdom}

\begin{abstract}
The Super Wide Angle Search for Planets (SuperWASP) is a whole-sky
high-cadence optical survey which has searched for exoplanetary
transit signatures since 2004.  Its archive contains long-term light
curves for $\sim$30 million 8--15~$V$ magnitude stars, making it a
valuable serendipitous resource for variable star research.  We have
concentrated on the evidence it provides for eclipsing binaries, in
particular those exhibiting orbital period variations, and have
developed custom tools to measure periods precisely and detect period
changes reliably.  Amongst our results are: a collection of 143
candidate contact or semi-detached eclipsing binaries near the
short-period limit in the main sequence binary period distribution; a
probable hierarchical triple exhibiting dramatic sinusoidal period
variations; a new doubly-eclipsing quintuple system; and new evidence
for period change or stability in 12 post-common-envelope eclipsing
binaries, which may support the existence of circumbinary planets in
such systems.  A large-scale search for period changes in $\sim$14000
SuperWASP eclipsing binary candidates also yields numerous examples of
sinusoidal period change, suggestive of tertiary companions, and may
allow us to constrain the frequency of triple systems amongst low-mass
stars.
\end{abstract}

\section{Introduction}
The Super Wide Angle Search for Planets: SuperWASP \citep{pollacco}
is, as its name indicates, primarily a sky survey searching for
exoplanetary transits.  However, several of its characteristics also
suit it well for the study of variable stars, especially eclipsing
binaries: it has an 8--15~$V$ magnitude range, a fairly long (7--9
years) time base, and a usefully high cadence (6--40 minutes).  Almost
whole-sky coverage is achieved, and the archive contains some 30
million light curves of which perhaps 1 million are significantly
variable.  The initial motivation for our study was to explore this
archival resource looking for evidence of orbital period variations in
eclipsing binaries; this was certainly achieved, and a number of
serendipitous discoveries of unusual systems were also made in the
course of the research programme.  Here we will outline some of our
analytical techniques and our main results of relevance to this
conference.

\section{Methods}
Techniques for detecting and quantifying the orbital periods and
period changes of eclipsing binary candidates from their SuperWASP
light curves were developed and improved over the course of the
project \citep{lohr,lohr13,lohr14b}.  The most reliable and precise
approach found for the large quantities of often noisy data available
here was to fold each full light curve on a range of trial periods and
select the period giving the least scatter of points about the mean
curve (a form of phase dispersion minimization): this did not require
any assumptions to be made about the underlying light curve shape.  In
addition to a value for the orbital period, this yielded a
phase-folded light curve which was used as a template to fit each
night of observations in the full curve.  The template was adjustable
in time, flux and amplitude, allowing optimal measurement of the times
of primary eclipse.  These observed times of minima could then be
compared with calculated times of minima (on the assumption of
constant period) to produce observed minus calculated ($O-C$)
diagrams: a standard tool for detecting and measuring period
variation.

\section{Results}

\subsection{Eclipsing binary candidates near the short-period limit}
We first focused attention on eclipsing binaries with very short
orbital periods, near the well-known but still mysterious limit in the
period distribution for main sequence systems, around 0.20~d
e.g.~\citet{paczynski}.  \citet{norton} had found 53 candidates with
$P<20000$~s in the SuperWASP archive; in \citet{lohr} some of the
periods of these were corrected, and period changes were detected in
several cases, demonstrating the feasibility of using this data set
for measuring orbital period variation.  Then, in \citet{lohr13}, the
set of such objects was increased to 143, nearly all previously
unpublished, and their periods were seen to tail off smoothly as one
end of the larger distribution, rather than exhibiting a sharp
cut-off: this may help to constrain possible explanations for the
short-period limit.  Several of these candidates have since been
confirmed spectroscopically as binaries, and their parameters
determined e.g. \citet{lohr14}; others turned out to be rare and
fascinating systems in their own right, as described in the next two
sections.

\subsection{A probable low-mass contact binary in a triple system}\label{triple}
1SWASP~J234401.81$-$212229.1 (J234401) showed an apparent contact
binary light curve, with primary eclipses only slightly deeper than
secondary ones, and an extremely short period of variation
(18461.64~s).  Over the first four years of archive data it exhibited
very dramatic period decrease \citep{lohr}, but subsequent
observations by SuperWASP and David Boyd supported a sinusoidal
variation in its $O-C$ diagram \citep{lohr13}, with modulating period
$\sim4.2$~y.  However, SALT spectra for the system did not give any
clear indications of binarity and seemed rather to indicate a single
mid-K star.  In \citet{lohr13b} we explored a range of possible models
to explain these apparently conflicting findings, and favoured a
triple system containing a low-mass (M+M) contact binary in a 4.2~y
orbit with a much hotter K star, which dominates the spectrum.
Recently, \citet{koen} has made multicolour observations and further
analysis of J234401, which support our preferred model: notably, the
system is bluest during eclipses (when a cool component is obscured),
and the light curve amplitude increases with increasing wavelength
(since the variation is produced by stars with spectra peaking in the
near-infrared).  This would appear to be only the second confirmed
contact system consisting of two M dwarfs.

\subsection{A doubly-eclipsing quintuple system}
1SWASP~J093010.78+533859.5 (J093010) also initially appeared to be a very
short-period contact binary ($P=19674.47$~s), but with significant
non-Gaussian data scatter below the main folded light curve;
additional eclipses were also visible in the full curve.  On refolding
it on a longer period (112798.90~s) a second \emph{detached} binary
was revealed, and we noted that Hipparcos had observed two point
sources at this location (TYC~3807-759-1 and TYC~3807-759-2) separated
by just 1\farcs88; in \citet{lohr13} we announced this as plausibly the
sixth-known doubly-eclipsing quadruple system.

\articlefiguretwo{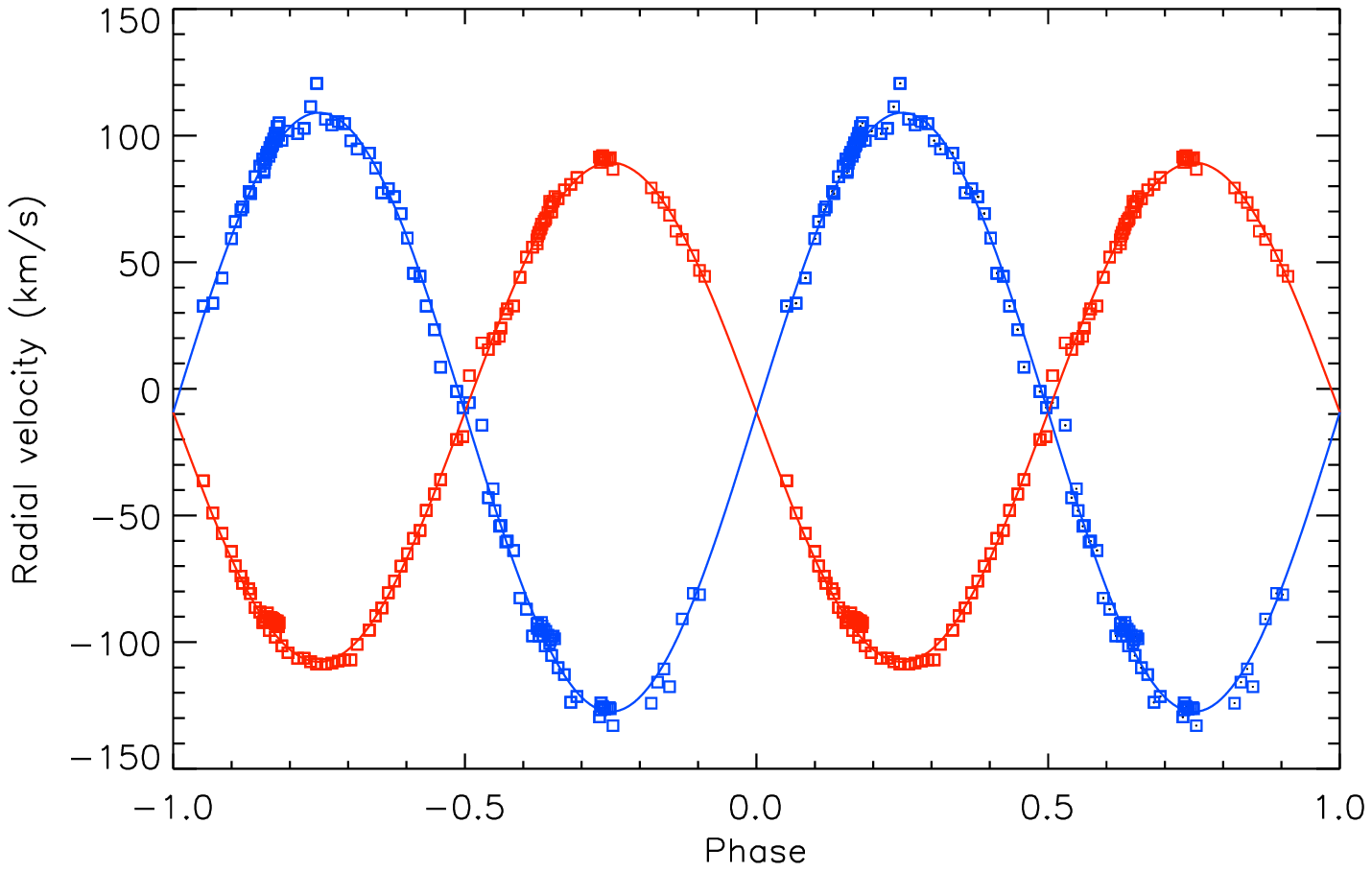}{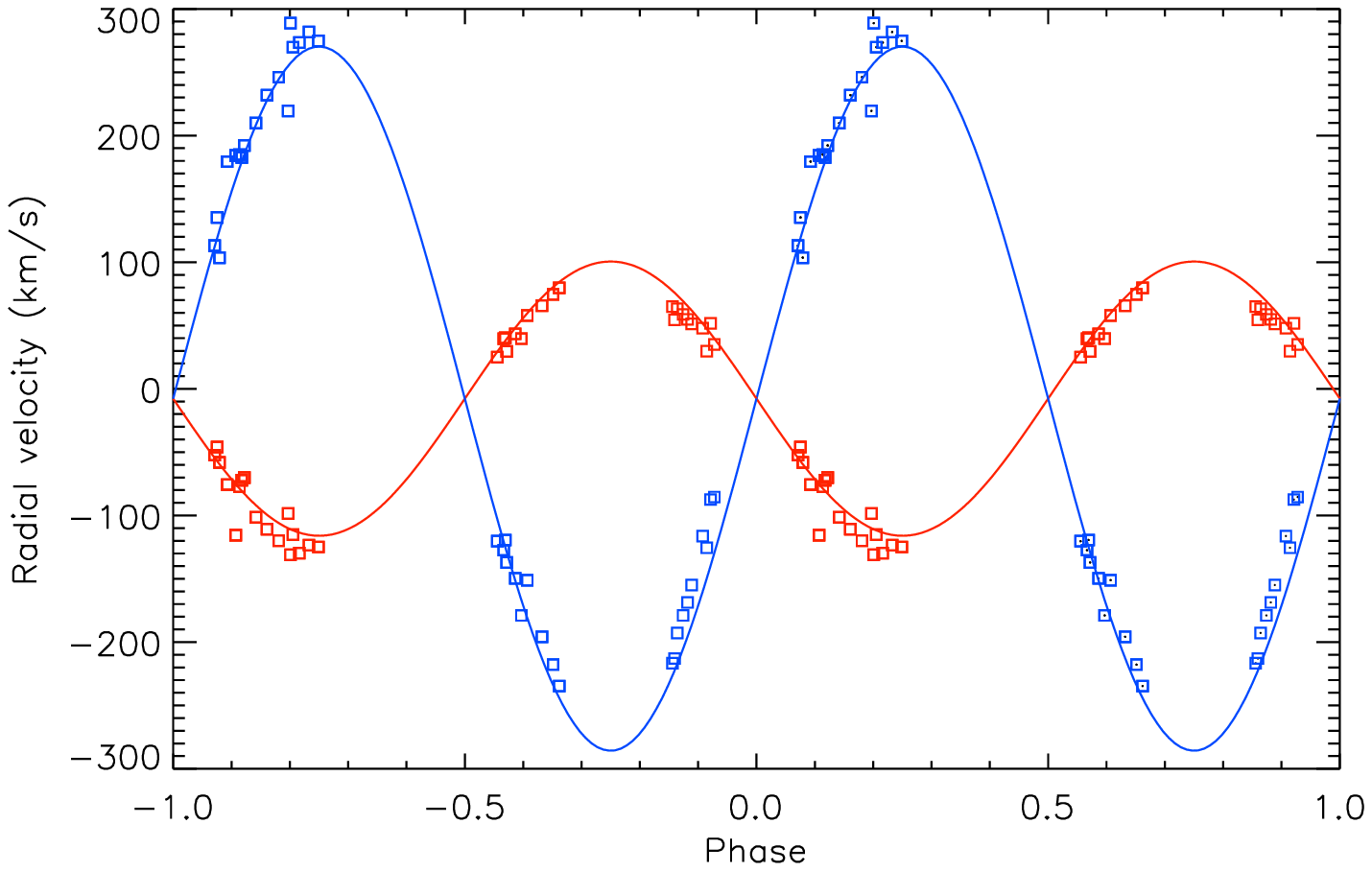}{rvcurves}{Radial
  velocity curves for J093010.  \emph{Left:} Detached system,
  combining our observations with those of \citet{koo}.  \emph{Right:}
  Contact system, based on our WHT observations.}

Since then, \citet{koo} obtained spectra for J093010, from which they
extracted radial velocities for the detached binary; they also noted
the presence of an extra, static set of spectral lines, which they
interpreted as a probable fifth star.  We have independently obtained
WHT spectra for each eclipsing system separately, and thus produced
radial velocity curves for both the detached and the contact systems
(Figure~\ref{rvcurves}), confirming each as a double-lined
spectroscopic and eclipsing binary \citep{lohr14c}.  Our spectra
confirm the presence of the fifth star, in close proximity to the
detached binary; moreover, all five stars exhibit a common systemic
velocity ($\sim-10$~km~s$^{-1}$) and compatible distance estimates
($\sim$50--60~pc), providing further support to their interpretation
as members of a gravitationally-bound quintuple.  Modelling of their
combining photometric and spectroscopic datasets also suggests an
angle of inclination $i\simeq89\deg$ for both binaries: it may be that
they lie in the same orbital plane and fragmented from a single
protostellar disk.

\subsection{Circumbinary planets in post-common-envelope systems?}
Another set of eclipsing binaries which we have investigated have
undergone common envelope evolution.  Such systems, of which the
prototype is HW~Vir, exhibit distinctive light curves with very
well-defined primary eclipses and strong reflection effects, allowing
precise determination of their times of light minima and tracking of
period changes.  Controversially, circumbinary planets have been
claimed in many such systems on this basis \citep{zoroschreib}.  In
\citet{lohr14b}, we reported our study of the 12 systems of this type
found in the SuperWASP archive, for which we measured hundreds of new
times of minima, some with uncertainties as small as a few seconds.
For HW~Vir, we found highly significant evidence of period increase,
which tallied perfectly with the two-planet model of
\citet{beuermann12b}; in NY~Vir, QS~Vir and NSVS~14256825 we found
less significant support for proposed period changes.  Also, there was
significant evidence of period change in ASAS~J102322$-$3737.0, which
had not previously been detected.  Whether or not circumbinary planets
are the cause, our findings strongly support the reality of period
variations in at least some eclipsing post-common-envelope systems.

\subsection{Multiplicity everywhere?}
Most recently, we have applied our analytical approach to $\sim14000$
SuperWASP eclipsing binary candidates identified in \citet{payne}.
Many cases of steady period increase and decrease (revealed by
quadratic fits to $O-C$ diagrams) were found for the three light
curve-defined classes of eclipsing system (EA, EB and EW);
unexpectedly, all the period change distributions were symmetrical
about zero, and very similar in shape.

\articlefigure{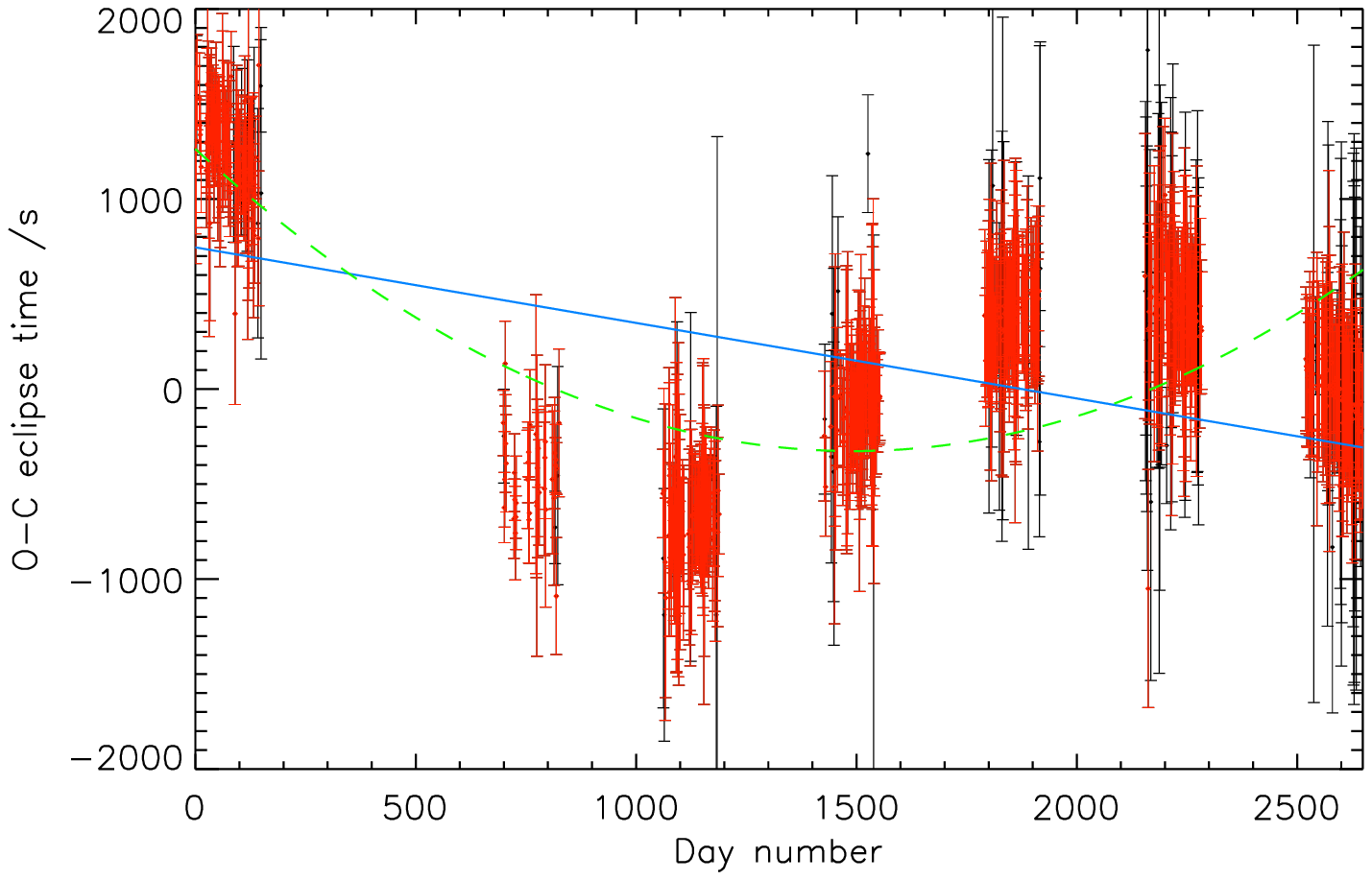}{sinoc}{$O-C$ diagram for a SuperWASP
  eclipsing binary candidate exhibiting sinusoidal variation.  Linear
  (solid line) and quadratic (dashed line) best fits are overplotted;
  neither matches the period variation well.}

A number of sinusoidally-varying $O-C$ diagrams were also seen
e.g. Figure~\ref{sinoc}, and in these may lie an explanation for the
symmetrical distributions of steady period changes \citep{lohr14c}.
The majority of apparently quadratic $O-C$ diagrams may in fact be
short sections of sinusoidal variations: being equally likely to come
from any part of a longer-term sinusoidal curve, they will be detected
as apparent steady period increases or decreases in equal numbers.  If
this is true, and if the majority of such sinusoidal period variations
are caused by undetected third bodies (as in Sect.\ \ref{triple}), we
can use these period change statistics to estimate the higher-order
multiplicity fraction: $\sim24$\% of SuperWASP binaries would be in
triples, a number which tallies well with recent estimates by other
researchers \citep{tokovinin1,tokovinin2,rappaport}.

\section{Conclusion}
The SuperWASP archive has proved to be a treasure trove of data on
eclipsing binaries, triples, and higher-order multiples.  We have been
able to uncover useful findings about specific sets of objects (very
short-period binaries, post-common-envelope systems) and discover rare
and informative individual systems (J234401, J093010).  We may even be
able to draw conclusions about the frequency of triple systems among
lower-mass stars in general.

\acknowledgements The WASP project is currently funded and operated by
Warwick University and Keele University, and was originally set up by
Queen's University Belfast, the Universities of Keele, St. Andrews and
Leicester, the Open University, the Isaac Newton Group, the Instituto
de Astrofisica de Canarias, the South African Astronomical Observatory
and by STFC.  This work was supported by the Science and Technology
Funding Council and the Open University.

\bibliography{lohrrefs}  

\begin{thebibliography}{}
\expandafter\ifx\csname natexlab\endcsname\relax\def\natexlab#1{#1}\fi
\expandafter\ifx\csname url\endcsname\relax
  \def\url#1{\texttt{#1}}\fi
\expandafter\ifx\csname urlprefix\endcsname\relax\def\urlprefix{URL }\fi
\providecommand{\eprint}[2][]{\url{#2}}

\bibitem[{Beuermann et~al.(2012)Beuermann, Dreizler, Hessman, \&
  Deller}]{beuermann12b}
Beuermann, K., Dreizler, S., Hessman, F.~V., \& Deller, J. 2012, A\&A, 543, 138

\bibitem[{Koen(2014)}]{koen}
Koen, C. 2014, MNRAS, 441, 3075

\bibitem[{Koo et~al.(2014)Koo, Lee, Lee, Kim, Lee, Hong, Lee, \& Rey}]{koo}
Koo, J.-R., Lee, J.~W., Lee, B.-C., Kim, S.-L., Lee, C.-U., Hong, K., Lee,
  D.-J., \& Rey, S.-C. 2014, AJ, 147, 104

\bibitem[{Lohr(2014)}]{lohr14c}
Lohr, M.~E. 2014, Ph.D. thesis, Open University

\bibitem[{Lohr et~al.(2014{\natexlab{a}})Lohr, Hodgkin, Norton, \&
  Kolb}]{lohr14}
Lohr, M.~E., Hodgkin, S.~T., Norton, A.~J., \& Kolb, U.~C. 2014{\natexlab{a}},
  A\&A, 563, A34

\bibitem[{Lohr et~al.(2014{\natexlab{b}})Lohr, Norton, Anderson, Cameron,
  Faedi, Haswell, Hellier, Hodgkin, Horne, Kolb, Maxted, Pollacco, Skillen,
  Smalley, West, \& Wheatley}]{lohr14b}
Lohr, M.~E., Norton, A.~J., Anderson, D.~R., Cameron, A.~C., Faedi, F.,
  Haswell, C.~A., Hellier, C., Hodgkin, S.~T., Horne, K., Kolb, U.~C., Maxted,
  P. F.~L., Pollacco, D., Skillen, I., Smalley, B., West, R.~G., \& Wheatley,
  P.~J. 2014{\natexlab{b}}, A\&A, 566, A128

\bibitem[{Lohr et~al.(2012)Lohr, Norton, Kolb, Anderson, Faedi, \& West}]{lohr}
Lohr, M.~E., Norton, A.~J., Kolb, U.~C., Anderson, D.~R., Faedi, F., \& West,
  R.~G. 2012, A\&A, 542, A124

\bibitem[{Lohr et~al.(2013{\natexlab{a}})Lohr, Norton, Kolb, \& Boyd}]{lohr13b}
Lohr, M.~E., Norton, A.~J., Kolb, U.~C., \& Boyd, D. R.~S. 2013{\natexlab{a}},
  A\&A, 558, A71

\bibitem[{Lohr et~al.(2013{\natexlab{b}})Lohr, Norton, Kolb, Maxted, Todd, \&
  West}]{lohr13}
Lohr, M.~E., Norton, A.~J., Kolb, U.~C., Maxted, P. F.~L., Todd, I., \& West,
  R.~G. 2013{\natexlab{b}}, A\&A, 549, A86

\bibitem[{Norton et~al.(2011)Norton, Payne, Evans, West, Wheatley, Anderson,
  Barros, Butters, Cameron, Christian, Enoch, Faedi, Haswell, Hellier, Holmes,
  Horne, Kane, Lister, Maxted, Parley, Pollacco, Simpson, Skillen, Smalley,
  Southworth, \& Street}]{norton}
Norton, A.~J., Payne, S.~G., Evans, T., West, R.~G., Wheatley, P.~J., Anderson,
  D.~R., Barros, S. C.~C., Butters, O.~W., Cameron, A.~C., Christian, D.~J.,
  Enoch, B., Faedi, F., Haswell, C.~A., Hellier, C., Holmes, S., Horne, K.~D.,
  Kane, S.~R., Lister, T.~A., Maxted, P. F.~L., Parley, N., Pollacco, D.,
  Simpson, E.~K., Skillen, I., Smalley, B., Southworth, J., \& Street, R.~A.
  2011, A\&A, 528, A90

\bibitem[{Paczy\'{n}ski et~al.(2006)Paczy\'{n}ski, Szczygie\l, Pilecki, \&
  Pojma\'{n}ski}]{paczynski}
Paczy\'{n}ski, B., Szczygie\l, D.~M., Pilecki, B., \& Pojma\'{n}ski, G. 2006,
  MNRAS, 368, 1311

\bibitem[{Payne(2013)}]{payne}
Payne, S.~G. 2013, Ph.D. thesis, Open University

\bibitem[{Pollacco et~al.(2006)Pollacco, Skillen, Cameron, Christian, Hellier,
  Irwin, Lister, Street, West, Anderson, Clarkson, Deeg, Enoch, Evans,
  Fitzsimmons, Haswell, Hodgkin, Horne, Kane, Keenan, Maxted, Norton, Osborne,
  Parley, Ryans, Smalley, Wheatley, \& Wilson}]{pollacco}
Pollacco, D.~L., Skillen, I., Cameron, A.~C., Christian, D.~J., Hellier, C.,
  Irwin, J., Lister, T.~A., Street, R.~A., West, R.~G., Anderson, D., Clarkson,
  W.~I., Deeg, H., Enoch, B., Evans, A., Fitzsimmons, A., Haswell, C.~A.,
  Hodgkin, S., Horne, K., Kane, S.~R., Keenan, F.~P., Maxted, P. F.~L., Norton,
  A.~J., Osborne, J., Parley, N.~R., Ryans, R. S.~I., Smalley, B., Wheatley,
  P.~J., \& Wilson, D.~M. 2006, PASP, 118, 1407

\bibitem[{Rappaport et~al.(2013)Rappaport, Deck, Levine, Borkovits, Carter,
  Mellah, Sanchis-Ojeda, \& Kalomeni}]{rappaport}
Rappaport, S., Deck, K., Levine, A., Borkovits, T., Carter, J., Mellah, I.~E.,
  Sanchis-Ojeda, R., \& Kalomeni, B. 2013, ApJ, 768, 33

\bibitem[{Tokovinin(2014{\natexlab{a}})}]{tokovinin1}
Tokovinin, A. 2014{\natexlab{a}}, AJ, 147, 86

\bibitem[{Tokovinin(2014{\natexlab{b}})}]{tokovinin2}
--- 2014{\natexlab{b}}, AJ, 147, 87

\bibitem[{Zorotovic \& Schreiber(2013)}]{zoroschreib}
Zorotovic, M., \& Schreiber, M.~R. 2013, A\&A, 549, A95

\end{thebibliography}

\end{document}